\documentclass[lettersize,journal]{IEEEtran}
\usepackage{amsmath,amsfonts}
\usepackage{array}
\usepackage[caption=false,font=normalsize,labelfont=sf,textfont=sf]{subfig}
\usepackage{textcomp}
\usepackage{stfloats}
\usepackage{url}
\usepackage{verbatim}
\usepackage{graphicx}
\usepackage{graphicx}
\usepackage[dvipsnames]{xcolor}
\usepackage{amssymb}
\usepackage{amsmath}
\usepackage{optidef}
\usepackage{array}
\usepackage{algpseudocode}
\usepackage{multirow}
\usepackage{algorithm}
\usepackage{cite}
\usepackage{csquotes}
\usepackage[hidelinks]{hyperref}
\usepackage{float}
\usepackage{booktabs}
\DeclareMathOperator*{\argmin}{arg\,min}

\def\BibTeX{{\rm B\kern-.05em{\sc i\kern-.025em b}\kern-.08em
    T\kern-.1667em\lower.7ex\hbox{E}\kern-.125emX}}
\usepackage{balance}
\begin{document}
\title{Anchor-Aided Multi-User Semantic Communication with Adaptive Decoders}
\author{Loc X. Nguyen, Phuong-Nam Tran, Trung Thanh Pham, Avi Deb Raha, Eui-Nam Huh,\\ Zhu Han,~\IEEEmembership{Fellow,~IEEE,} and Choong Seon Hong,~\IEEEmembership{Fellow,~IEEE}
\thanks{Loc X. Nguyen, Phuong-Nam Tran, Trung Thanh Pham, Avi Deb Raha, Eui-Nam Huh, Choong Seon Hong are with the School of Computing, Kyung Hee University, Yongin-si, Gyeonggi-do 17104, Rep. of Korea, e-mail: \{xuanloc088, tpnam0901, trungpt, avi, johnhuh, cshong\}@khu.ac.kr.}

\thanks{Zhu Han is with the Electrical and Computer Engineering Department, University of Houston, Houston, TX 77004, and also with the Department of Computer Science and Engineering, Kyung Hee University, Yongin-si, Gyeonggi-do 17104, Rep. of Korea, e-mail:{\{hanzhu22\}}@gmail.com}
}

\markboth{IEEE Transactions on Network Science and Engineering, May~2026}%
{How to Use the IEEEtran \LaTeX \ Templates}

\maketitle

\begin{abstract}

Semantic communication (SemCom) is accelerating its momentum to catch up with the massive increase in users' demands in both quantity and quality, with the assistance of advanced deep learning (DL) techniques. Specifically, SemCom can actively embed the semantic meaning of the data into the transmission process, while eliminating statistical redundancy to preserve bandwidth resources for other users. Therefore, the transmitter encodes the message in the most concise way, while the receiver tries to interpret the message with the DL model and its knowledge of the transmitter’s intended meaning. Most existing works only consider one transmitter and one receiver, which limits their ability to address the diversity in users' models and capabilities. Therefore, in this paper, we propose a multi-user semantic communication system where each user is equipped with a distinct DL-based joint source-channel decoder architecture, reflecting the diversity in computing capacity. The challenging issue with the proposed system is the catastrophic forgetting property of neural networks, where the DL-based encoder fails to encode the data for the previous user when being trained with a new user. To address this, we propose an anchor decoder with an architecture that is symmetric to the encoder. The symmetric decoder has the same computational capacity as the encoder, providing feedback that aligns with the encoder's extraction capabilities and enhances optimization efficiency. The parameters of the optimized encoder are then frozen and used to train decoders for various users, aligning them with the encoder outputs. Finally, we conduct a series of simulation experiments to validate the proposed framework against other benchmarks.
\end{abstract}

\begin{IEEEkeywords}
Semantic Communication, Diverse Computing Capacity, Deep Joint Source-Channel Coding, Multi-users, Cross-architecture, Internet of Things devices, Catastrophic forgetting, Heterogeneous devices.
\end{IEEEkeywords}

\section{Introduction}

\IEEEPARstart{T}{here} are three main factors that have caused significant challenges for wireless communication systems in recent years: the explosive growth in data traffic, the rapid expansion of connected devices, and the demanding low-latency needs of new applications, such as real-time gaming and autonomous driving. Researchers have explored ways to address these problems, including integrating sensing technologies to enhance hardware efficiency and employing sophisticated network management techniques. Nevertheless, conventional communication techniques are approaching the Shannon capacity limit of the physical layer~\cite{luo2022semantic}, which hinders further advancements in transmission speeds. As a result, research and business investment are now heavily focused on creating more effective wireless communication paradigms. One notable approach in recent years is the semantic communication system, which addresses the communication bottleneck by directly reducing the amount of data transmitted through the wireless channel~\cite{9959884,nguyen2025contemporary,9955312}. Specifically, the transmitter will actively determine the importance of the information within the message to the receiver, and then remove the redundant information from the transmission process to preserve the communication bandwidth. On the other side, the semantic communication receiver interprets the message meaning with its knowledge of the transmitter and its decoders. By prioritizing the qualitative over the quantitative, the semantic communication paradigm requires significantly lower-bandwidth channels to communicate, thus preserving resources for other devices.

The concept of semantic communication was first introduced by Claude Shannon and Warren Weaver in the 1950s~\cite{shannon1949mathematical}, where the objective extends beyond accurate bit-level reconstruction in conventional communication systems to the successful interpretation of the underlying message meaning at the receiver. Despite this early vision, semantic communication had not achieved compelling performance or practical deployment. This limitation primarily arises from the lack of effective methodologies for extracting semantic information from raw data and encoding it into compact representations suitable for transmission. Recent advances in deep learning (DL) have significantly changed this landscape. In particular, DL-based models enable end-to-end learning of semantic representations, allowing the encoder to jointly perform feature extraction and compression in a data-driven manner~\cite{bourtsoulatze2019deep}. This paradigm differs from traditional communication systems, where source and channel coding are designed and optimized separately, often leading to suboptimal performance under complex channel conditions.

Motivated by these advantages, numerous works have adopted the DL-based joint source-channel coding (D-JSCC) as a practical realization of semantic communication, with the application spanning the text~\cite{farsad2018deeptext,9398576,10000901}, image~\cite{9953076,9066966,10094735}, and speech~\cite{9953316,10038754,9450827} modality. By jointly optimizing the source representation and channel protection in an end-to-end manner, D-JSCC extracts task-relevant information directly from the data and embeds the semantic representation in a form that is intrinsically robust to channel noise and fading.  A particularly notable advantage of this paradigm is its ability to overcome the so-called \emph{cliff effect} of conventional separation-based systems, in which reconstruction quality collapses abruptly once the channel signal-to-noise ratio falls below the threshold dictated by the chosen code rate; D-JSCC, in contrast, exhibits graceful degradation across a wide range of channel conditions. Beyond the D-JSCC-based direction, there exist two other directions to pursue the realization of a semantic communication system: Generative AI-based SemCom and Theory-of-mind-based SemCom, both of which lean toward task-oriented objectives rather than the faithful data reconstruction. The present paper restricts its attention to the D-JSCC direction and addresses one of its key open challenges through an adaptive training strategy.

Progress along the D-JSCC direction has been closely tied to advances in deep learning architectures, with each successive generation of models adopted in turn as the semantic encoder in order to extract richer task-relevant features from the source. Early designs were built on convolutional neural networks~\cite{bourtsoulatze2019deep}, subsequently augmented with channel-wise attention modules to provide signal-to-noise-ratio adaptivity~\cite{xu2021wireless}, and most recently equipped with Transformer backbones~\cite{10854360} and their variants tailored to wireless image transmission~\cite{10094735}. Despite this rapid architectural progress, the overwhelming majority of D-JSCC studies remain confined to single-user settings, and only a handful have considered multi-user scenarios~\cite{9830752,10972177}. Among these, an even smaller subset has acknowledged the heterogeneity in users' computing capacities~\cite{10755087}; and even those works typically assume that every user employs an identical DL backbone, differing only in network depth or width. Although such assumptions ease optimization and yield favorable convergence, they significantly compromise both the scalability of the resulting system and its fidelity to real-world deployments, in which device populations are architecturally diverse by necessity rather than by choice.

The case for relaxing this homogeneity assumption is further reinforced by the explosive growth in connected devices and the breadth of intelligent services that modern communication systems must support, ranging from fully automated environments such as smart factories and smart cities to continuous human connectivity across heterogeneous applications and contexts. D-JSCC-enabled devices have emerged as a compelling building block for these scenarios, offering reliable transmission even under harsh channel conditions~\cite{9252948}. In practice, however, such devices are produced by different manufacturers and under different design objectives, resulting in substantial heterogeneity across their technical specifications, including power budget, computational capability, and wireless communication performance~\cite{9746463}. Existing D-JSCC SemCom designs nevertheless deploy a single identical model across all devices, which is increasingly difficult to justify in light of this heterogeneity. A more realistic design instead equips each device with a distinct deep learning model for the JSCC module, tailored to its on-device resources, while sharing a common encoder at the base station.

In this work, we investigate a multi-user semantic communication system, in which the users are architecturally heterogeneous, reflecting the asynchronous and resource-constrained nature of practical devices in a real-world network. Specifically, we consider downlink transmission from a base station (BS) to multiple users, each equipped with a distinct DL-based decoder architecture tailored to its computational capability. Such a cross-architecture poses a critical challenge for training DL models in the network: the shared BS encoder received different updated directions from clients in the network, which leads to the \textit{catastrophic forgetting}, as its parameters are continuously optimized using the gradients originating from heterogeneous decoder models. To address this challenge, we propose an anchor-aided two-stage training framework that decouples encoder optimization from decoder adaptation, stabilizing the learned representations while preserving the flexibility to serve an arbitrary number of architecturally diverse users. The main contributions are summarized as follows:
\begin{itemize}
    \item \textbf{Cross-architecture multi-user SemCom:} We consider a multi-user D-JSCC semantic communication system where each user is equipped with a decoder based on a different DL architecture. This design better reflects real-world scenarios, where devices vary in computational capability and their underlying model structures. Unlike most existing works that rely on a shared backbone with only minor modifications, our setting introduces genuine architectural diversity. As a result, it brings to light an important yet largely overlooked challenge: training instability arising from conflicting gradient updates when heterogeneous decoders interact with a shared encoder.
    
    \item \textbf{Anchor-aided two-stage training framework:} We propose a two-stage training strategy to address the gradient conflict problem, which is the root cause of forgetting at the BS encoder. In the first stage, the encoder is trained together with a decoder that shares a symmetric architecture, which promotes stable learning and allows the encoder to fully capture semantic representations. In the second stage, the encoder parameters are fixed and treated as an anchor, while user-specific decoders are trained to adapt to their output. This design enables efficient adaptation across users without degrading the representations learned by the encoder.

    \item \textbf{Robustness and scalability evaluations:} Extensive simulations over both additive white Gaussian noise (AWGN) and Rayleigh fading channels demonstrate the effectiveness of the proposed framework compared to conventional iterative and joint training schemes. In particular, the proposed approach effectively mitigates the encoder forgetting problem and enables the system to scale efficiently with an increasing number of users.
\end{itemize}
The remainder of the paper will be organized as follows. We will provide a comprehensive discussion of the related works from the foundation theory of semantic communication, to current modern works and existing direction in Section \ref{Related}. Then, Section \ref{System} describes a general system model of the semantic communication and its common end-to-end training. Our proposed solution will be demonstrated in-depth in Section \ref{Proposed}. We present the simulation results and, based on them, draw conclusions in Sections \ref{Performance} and \ref{Conclusion}, respectively.

\section{Related works}\label{Related}

This section provides an overview of the research areas most relevant to our work. We first review the paradigm of semantic communication and its main directions of development. Next, we discuss the current trends and advancements of D-JSCC methods for image transmission, followed by their extensions to multi-user scenarios, which motivate our problem formulation. We then examine related studies in continual learning, heterogeneous model design, and decoupled training. Finally, we summarize the key distinctions between our approach and the most closely related prior work.

\subsection{Foundations and Taxonomy of Semantic Communication}
The notion that the ultimate goal of communication need not be exact bit reproduction predates modern learning by several decades. Weaver, in his commentary accompanying Shannon's mathematical theory of communication~\cite{shannon1949mathematical}, explicitly distinguished three levels of communication problems: 
\begin{itemize}
    \item Technical level: How accurately can symbols be transmitted?
    \item Semantic level: How precisely do the transmitted symbols convey the desired meaning?
    \item Effectiveness level: how does the received meaning affect conduct? 
\end{itemize}
And the classical information theory addresses only the first level. The semantic level was given its first formal treatment by Bar-Hillel and Carnap~\cite{carnap1952outline}, who introduced a logical-probabilistic measure of semantic content based on the set of possible worlds excluded by a proposition. Although these early efforts did not yield engineerable systems, they established the conceptual scaffolding that contemporary semantic communication seeks to fill with DL. Modern SemCom has been propelled by DL's ability to extract task-relevant features from raw signals, and the literature has converged on three principal technical directions, each of which can be further organized along a task-oriented versus reconstruction-oriented axis~\cite{nguyen2025contemporary,9955312}.

The first direction, \textbf{Deep Joint Source-Channel Coding}, replaces the classical separation of source and channel coding with a single end-to-end-trained neural network that maps source symbols directly to the channel output and back. The seminal works of Bourtsoulatze \textit{et al.}~\cite{bourtsoulatze2019deep} established that a convolution autoencoder trained jointly with differentiable channel layers can outperform the separated JPEG/LDPC pipelines, especially under the low signal-to-noise ratio (SNR). Under conventional communication, the pipeline experiences an abrupt collapse in reconstruction quality when channel conditions fall below a threshold, at which the modules cannot perform bit correction effectively; this phenomenon is frequently referred to as the \textit{cliff effect}. On the other hand, the D-JSCC adaptively adjusts its coding strategy according to the channel condition~\cite{9066966,10531097}, therefore degrades gracefully. The direction has been expanded across modalities, such as text, image, speech, and video, and has been verified for its advantages in these modalities. The developments of D-JSCC are closely associated with the advancement in DL architecture, since the complex models can efficiently identify and extract the semantic features from raw data. 

The second direction, \textbf{Generative-AI-based SemCom} Recent works \cite{grassucci2023generative,11263916,10447237,11039171,11493754,10646587} have explored generative AI-based semantic communication, shifting the focus from exact data reconstruction to conveying semantically equivalent information. In this paradigm, the transmitter aims to deliver the underlying meaning of the data rather than the raw signal itself. For example, instead of transmitting an image directly, the transmitter may encode its semantic content in textual form, which typically requires significantly less bandwidth than image-based transmission~\cite{10447237}. At the receiver, this textual description serves as a condition for a generative model to reconstruct an image that preserves the semantic meaning of the original input. Building on this idea, generative AI models can be deployed at both the transmitter and receiver, enabling flexible cross-modal communication~\cite{11321267} and supporting multi-modal scenarios due to their strong generative and representation capabilities. As its name suggests, the approach is beneficial for the advancement of generative models such as diffusion~\cite{rombach2022high} and generative adversarial nets~\cite{goodfellow2014generative}. 

The third direction, \textbf{Theory-of-Mind-based SemCom}~\cite{10054510,10433140,10272264} goes one step further and treats communication devices as agents with cognitive entities, whose shared language emerges from contextual reasoning and repeated interaction. In particular, \cite{10054510} demonstrates a scenario in which the speaker and listener iteratively communicate to build a common language using the semantic triangle: observation-concept-symbol space. Contextual reasoning is used by the transmitter to predict the receiver's inferential knowledge when receiving the set of symbols, thereby determining the optimal set of symbols to convey the message at minimum cost. Throughout the iterations, knowledge on both sides of the communication evolves, gradually forming a concise language that mirrors human behavior~\cite{10272264}. However, the effectiveness of this approach strongly depends on the design and scalability of the underlying knowledge base, including the representation of inference rules, the efficiency of knowledge storage and update mechanisms, and the ability to generalize across diverse communication scenarios.

Among these directions, D-JSCC has emerged as the dominant paradigm in the SemCom literature, commanding the largest share of research attention owing to its empirical performance gains, and practical feasibility.
Crucially, D-JSCC has been validated beyond simulation: Yoo \textit{et al.}~\cite{yoo2022real} demonstrated its viability on a real-world hardware testbed, confirming that the performance advantages observed in theoretical studies translate to practical deployment environments. Motivated by this breadth of evidence, our paper builds on and extends the D-JSCC paradigm to address the heterogeneity of the DL architecture of communication participants.

\subsection{Deep Joint Source-Channel Coding for Image Transmission}

Within the D-JSCC family, image transmission has received the most sustained attention since they are bandwidth-hungry compared with text and speech, and can be extended to video if we consider videos as a set of image frames. We review this body of work along three axes that are directly relevant to our contribution: architecture evolution, channel and rate adaptivity, and digital/system integration.

\subsubsection{Architecture Evolution}

In the early stages, the D-JSCC system~\cite{bourtsoulatze2019deep} used purely convolutional encoders and decoders, with the channel modeled as a non-trainable additive Gaussian or Rayleigh layer inserted between them. Building on this work, authors in~\cite{9066966} noted the importance of embedding channel feedback into the encoding and decoding processes, thereby providing a complete view for the system to achieve robust performance with respect to the dynamics of channel noise. Approaching the channel feedback from a different perspective, \cite{xu2021wireless} considered integrating the SNR value of the wireless environment into the channel encoder/decoder, which provides the range of potential noise and enables the neural network to mitigate its effects. The work~\cite{10094735} acknowledged the capability of attention mechanisms in semantic feature extraction; however, this advantage comes at the cost of high computational complexity. Therefore, they leveraged the Swin Transformer to reduce complexity by computing attention within local windows rather than across the entire image. In parallel, the work~\cite{9791398} combined nonlinear transform coding with the Transformer-based D-JSCC to achieve an adaptive compression rate. Recently, Mamba Vision~\cite{hatamizadeh2025mambavision} has gained attention from the research community as a potential replacement for Transformer-based architectures, and has also been adopted in semantic communication~\cite{10975284} for image transmission. Collectively, these works illustrate that the evolution of D-JSCC is deeply intertwined with broader advances in deep learning architecture, from convolutional networks and channel-adaptive mechanisms to attention-based models and state-space architectures. 

\subsubsection{Channel and Rate Adaptive}

A practical D-JSCC system is required to cope with two factors of variability: the dynamics of channel quality and the available bandwidth~\cite{10388062}. The SNR-adaptive design~\cite{11249452} injects an estimated SNR value into the intermediate feature maps so that a single network can operate in a wide range of conditions. The robustness of the SemCom system not only depends on the channel conditions but also on the current task. The work~\cite{11145928} employed reinforcement learning to adapt the compression rate based on the task and channel conditions. On the other hand, the authors in \cite{10423076} proposed a mechanism to adjust the compression rate based on the network state, thereby actively reducing communication overhead in SemCom. In parallel, the work~\cite{10981842} proposed bandwidth adaptation by prioritizing transmitting the first $C$ semantic features. The information bottleneck theory is considered to design a rate-adaptive mechanism, whose objective is to balance the rate, semantic-level distortion, and semantic entropy~\cite{11219364}. In summary, these works underscore that adaptability across channel conditions, bandwidth constraints, and task requirements is a central design principle for practical systems.

\subsubsection{Digital Integration and System-level Extensions}
Most D-JSCC systems output continuous-valued symbols, which are incompatible with modern digital communication. Hu \textit{et al.}~\cite{10101778}  proposed vector quantization for the output signal, then leveraged a shared codebook between communication devices to further combat the semantic noise. Park \textit{et al.}~\cite{10584091} addressed this by introducing learned constellation mappings and uncertainty-aware demodulation, producing a fully digital D-JSCC pipeline with adjustable modulation order. The work~\cite{10288558} extended the system to tackle multiple access by jointly training D-JSCC pipeline with orthogonal signal modulation to prevent interference. In addition, an in-band full-duplex semantic communication was proposed in \cite{11123575}, where the authors integrated D-JSCC with conventional radio frequency communication to preserve transmission resources and guarantee the integrity and reliability of semantic information. These system-level extensions are complementary to our contribution, which addresses the upstream problems.

\subsection{Multi-user Semantic Communication}

Although multi-user wireless systems are the rule rather than the exception in practice, multi-user SemCom remains comparatively underexplored. In this subsection, we organize the existing literature into overlapping strands and identify the gaps our paper addresses.

\subsubsection{Broadcast and one-to-many D-JSCC}

Most of the existing works on multi-user semantic communication focused on the simplest setting, which is downlink broadcast: a single transmitter serves multiple receivers that share an architecturally identical decoder. An interesting work was proposed in~\cite{9885016}, where the authors acknowledge that semantic meaning varies across users and capture the semantic recognizer using DistilBERT. Luo et al.~\cite{10981842} proposed an adaptive D-JSCC scheme that re-weights latent channels on a per-receiver basis to reflect heterogeneous bandwidth allocations, but the underlying decoder topology is shared. In~\cite{11045090}, the author highlighted the lack of style features in multi-user semantic communication systems and thus proposed a dual-branch semantic encoder/decoder to extract both semantic content and style features, thereby improving reconstruction quality. The authors in \cite{11112686} focused on redesigning the DL architecture for the joint source-channel coding module and optimizing the formulated utility maximization with the soft actor-critic network for achieving higher reliability. Unlike the other works, \cite{10622730} considered the interference in multi-user signal reception and tackled it by proposing a diffusion-based channel enhancer that effectively eliminates interference from other users.  In \cite{11143317}, the authors considered the scenario where multiple users report the current task to the roadside unit, and thus, it changes the encoding strategy to be suitable for the task demand. These works establish that one-to-many SemCom is feasible but do not address the problem that arises when receivers also differ in compute capacity and DL backbone.
\subsubsection{Multiple Access and Interference}

A second-strand handle for uplink and shared-spectrum multiple access. Here, \cite{10225385} integrated D-JSCC with non-orthogonal multiple access (NOMA), letting the receiver jointly decode superposed semantic streams. The work in~\cite{10158994} illustrated that opportunistic switching between SemCom and BitCom can improve resource efficiency and user performance under the NOMA framework. The authors in~\cite{10505144} proposed Orthogonal-Model Division Multiple Access with the objective to migrate the anti-interference capability. Inspired by the orthogonal frequency-division multiplexing (OFDM) communication architecture, a semantically important-aware reordering-enhanced system is developed, along with channel estimation, to reduce the channel estimation interpolation error in~\cite{10843133}. 
\begin{table*}[!t]
    \centering
    \caption{Comparison of representative semantic communication frameworks with respect to multi-user support architectural heterogeneity. ``Heterogeneous architecture family'' indicates whether the framework supports fundamentally different decoder architectures rather than only scaled variants of the same model family.}
    \label{tab:positioning}
    \renewcommand{\arraystretch}{1.3}
    \begin{tabular}{lcccc}
        \toprule
        \textbf{Reference} & \textbf{Multi-user} & \textbf{Heterog.\ arch.\ family} 
        & \textbf{Shared encoder} & \textbf{Forgetting addressed} \\
        \midrule
        Bourtsoulatze \textit{et al.}~\cite{bourtsoulatze2019deep} 
            & No  & N/A             & N/A       & No \\
        Xu \textit{et al.}~\cite{xu2021wireless}                    
            & No  & N/A             & N/A       & No \\
        Yang \textit{et al.} (WITT)~\cite{10094735}             
            & No  & N/A             & N/A       & No \\
        Li \textit{et al.} (NOMA)~\cite{10225385}                 
            & Yes & No (identical)  & Yes       & No \\
        Xie \textit{et al.}~\cite{9830752}                      
            & Yes & No (identical)  & Yes       & No \\
        Nguyen \textit{et al.}~\cite{10423076}               
            & Yes & No (depth only) & Yes       & No \\
        Nguyen \textit{et al.} (KD)~\cite{10755087}             
            & Yes & No (size only)  & Yes       & Implicit (KD) \\
        Song \textit{et al.}~\cite{11045090}               
            & Yes & No (identical)  & Yes       & No \\
        Luo \textit{et al.}~\cite{10981842}                  
            & Yes & No (identical)  & Yes       & No \\
        Zeng \textit{et al.}~\cite{10622730}                
            & Yes & No (identical)  & Yes       & No \\
        Tan \textit{et al.} (Fed)~\cite{10729265}           
            & Yes & No (identical)  & No (FL)   & No \\
        \midrule
        \textbf{This work} 
            & \textbf{Yes} & \textbf{Yes (4 families)} 
            & \textbf{Yes} & \textbf{Yes (Theorems 1--3)} \\
        \bottomrule
    \end{tabular}
\end{table*}
\subsubsection{The Remaining Gap}
Across the aforementioned works, none of the existing multi-user SemCom works simultaneously (a) employs distinct DL backbone families across decoders, (b) shares a single encoder across this heterogeneous population, and (c) explicitly addresses the catastrophic forgetting that arises when the encoder is updated in turn by gradients originating from these heterogeneous backbones. To our knowledge, this combination has not been studied before; the present paper is the first to examine it.

\subsection{Catastrophic Forgetting and Continual Learning}

Catastrophic forgetting is a long-standing phenomenon in neural networks~\cite{kirkpatrick2017overcoming} and typically occurs when the neural network is trained sequentially on tasks $[T_{1},T_{2},\ldots,T_{m}]$, the parameters that minimize the loss on $T_{m}$ tend to drift way from regions that were good for $[T_{1}, \ldots, T_{m-1}]$, degrading earlier performance. Our setting is structurally similar: the BS encoder plays the role of the shared body, and the decoder of each user plays the role of a separate task. Theorem 1 in Section~\ref{theorem} is in fact a restatement, in our notation, of the gradient-conflict view of forgetting that underpins much of the modern continual-learning literature~\cite{rolnick2019experience}. The continual-learning methods are conventionally grouped into three families. 
\begin{itemize}
    \item Regularization-based methods add a penalty to the loss that discourages parameters from moving away from values that were important for early tasks. Elastic Weight Consolidation~\cite{kirkpatrick2017overcoming} reduced the learning rate for specific weights according to their significance for previously encountered tasks, while~\cite{zenke2017continual} proposed Synaptic Intelligence to record both past and current parameter values and maintain an online estimate of the synapse’s importance in addressing previously encountered tasks.
    \item Replay-based methods keep a buffer of past examples/synthetic ones, and replay them when training on new tasks. Experience Replay~\cite{rolnick2019experience} and Gradient Episodic Memory~\cite{NIPS2017_f8752278} aim to ensure the gradient on the new task does not increase the loss on stored examples.
    \item Architectural/Parameter-isolation methods allocate disjoint parameter subsets to different tasks and either freeze old subsets or grow the network. PackNet~\cite{mallya2018packnet} leverages redundancies in large deep networks to release parameters, which are subsequently utilized for learning new tasks through iterative pruning and network retraining.
\end{itemize}
\textit{Why none of these is a drop-in solution?} All three families assume a single learning agent facing a stream of tasks. In multi-user SemCom, in contrast, (i) the user population is fixed and known, (ii) the training data is the same across users (only the channel realization and decoder architecture differ), and (iii) the bottleneck is a shared component (the encoder) rather than the user-side networks. Our framework exploits these structural properties to obtain a particularly simple decoupling: train the encoder once with a strong, symmetric partner, then freeze it. This is the simplest possible parameter-isolation scheme, and Theorems 2 and 3 formalize why it suffices.

\subsection{Positioning of This Work}

In summary, the present paper occupies a previously unaddressed corner of the multi-user SemCom design space, defined by three properties simultaneously: (P1) the user population is architecturally heterogeneous across distinct DL families, not merely along width or depth; (P2) a single base-station encoder is shared across this population; and (P3) the training procedure must explicitly avoid the catastrophic forgetting that gradient flows from heterogeneous decoders would otherwise induce in the shared encoder. To make this positioning concrete, Table~\ref{tab:positioning} summarizes how representative prior multi-user SemCom works compared on these axes. Compared with prior multi-user works~\cite{10423076,10225385,9830752,10972177,10755087,11045090,10433140,10981842,10622730,11143317,11404244}, the proposed framework is distinguished by its explicit treatment of architectural heterogeneity as the source of forgetting and its anchor-based decoupling as a principled remedy.

\begin{figure*}[t]
    \centering
    \includegraphics[width=0.77\textwidth]{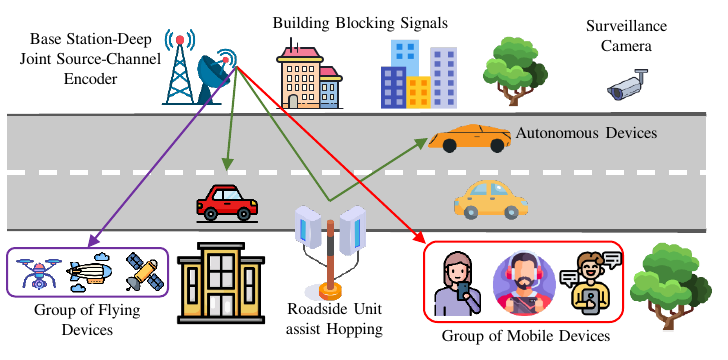}
        \caption{The proposed semantic communication system for various types of semantic users, including mobile devices, autonomous cars, and unmanned aerial vehicles. Each group possesses a distinct DL architecture, and users within the group can have a different number of layers.}
    \label{system_model_fig}
\end{figure*}

\section{System Model and Problem Formulation}\label{System}

\subsection{Overview of System Model}

In Fig.~\ref{system_model_fig}, we provide a detailed visualization of our proposed system, where the BS conducts the downlink communication process to users by using \textcolor{black}{the DL-based joint source-channel encoder for the selective encoding of important features of the message, therefore reducing the communication bandwidth and latency}. \textcolor{black}{On the other hand,} the semantic communication users span a wide range from mobile users, unmanned aerial vehicles, to autonomous cars, whose devices are equipped with a distinct DL architecture for the joint source-channel decoder. It is worth noticing that the devices from the same group can still be exposed to the difference in communication model, for example, different brands of mobile phones, or even the same brand but different in generation. \textcolor{black}{In addition, each group of clients has different priorities, and they also affect the deployment of DL architecture for the semantic communication, as shown in Table~\ref{requirements}}. Therefore, it is essential to consider this factor in the semantic communication system and reflect a real-world scenario. 

\begin{table}[ht]
\centering
\caption{\textcolor{black}{Different characteristics for deploying different DL models}}

\begin{tabular}{|c|c|c|c|c|}
\hline
\multirow{2}{*}{\textbf{Devices}} & \multicolumn{4}{c|}{\textbf{Requirements}} \\ \cline{2-5}
 & \textbf{Latency} & \textbf{Reliability} & \textbf{\begin{tabular}[c]{@{}c@{}}Energy \\ Available\end{tabular}} & \textbf{\begin{tabular}[c]{@{}c@{}}Computation \\ Capacity\end{tabular}} \\ \hline
\textbf{UAV} & \begin{tabular}[c]{@{}c@{}}Moderate\\ Low\end{tabular} & High & Low & Low \\ \hline
\textbf{Mobile} & Moderate & Medium & Low & \begin{tabular}[c]{@{}c@{}}Moderate\\ Low\end{tabular} \\ \hline
\textbf{\begin{tabular}[c]{@{}c@{}}Autonomous \\ Vehicles\end{tabular}} & Ultra low & Ultra-high & High & High \\ \hline
\textbf{\begin{tabular}[c]{@{}c@{}}Road Side \\ Unit\end{tabular}} & Moderate & High & Unlimited & Moderate \\ \hline
\end{tabular}
\label{requirements}
\end{table}
\color{black}
\subsection{Deep Joint Source-Channel Coding for BS and Users}
In the paper, we particularly focus on the transmission for image modality. We first denote the image notation as $I \in \mathbb{R}^{n}$, where $n$ represents the image size, calculated by multiplying the height, width, and number of color channels: $H \times W \times C$. Differentiating itself from conventional communication, the deep joint source-channel encoding approach of semantic communication optimizes both the source and channel modules simultaneously with its deep learning model. This particular property not only enables communication to achieve global optimality but also embeds the semantic meaning of the data into the transmission process, where the semantic encoder model learns how the different channel noise levels affect the transmitted signal. Therefore, the DL-based encoder adaptively adjusts the coding to combat the environment noise and achieves channel noise robustness. This encoding process is denoted as follows: 
\begin{equation}
    X_{I}= E_{\alpha}(I,\Gamma), \forall I \in \mathbb{R}^{n}, \forall \Gamma \in \mathrm{SNR},\label{EncodingEquation}
\end{equation}
where $E_{\alpha}(\cdot)$ denotes the joint source-channel encoder of the BS with the corresponding parameters $\alpha$. $\Gamma$ indicates the SNR, where its value can span a wide range due to the dynamic nature of wireless systems. As presented in the above equation, the SNR value is embedded into the encoding process of the BS to provide the channel condition and thus obtain an adaptive encoding strategy. The encoded signal $X_{I} \in \mathbb{R}^{k}$ has the size of $k$ and its values are significantly lower than the size of the original image $n$, and we denote the compression rate as $k/n$. It is obvious that our target is to obtain a low compression rate and thus secure a highly efficient communication, while maintaining the quality of the reconstruction image. However, the trade-off between compression rate and quality is inevitable, where we aim to find a sweet spot that balances communication cost and image quality. On the other hand, the semantic communication users are also equipped with a unique DL-based joint source-channel decoder to interpret the received signal. The encoded signal from the BS is transmitted over a wireless environment, where it is exposed to both noise and fading as described in the following equation:
\begin{equation}
    Y_{I,k}=X_{I,k}H_{k}+N_{k}, \forall N_{k} \in \mathcal{N}(0,\sigma^{2}\boldsymbol{I}),
\end{equation}
where $Y_{I,k}$ and $H_{k}$ are the received signal for the image $I$ at user $k$ and the fading coefficient between the BS and the receiver $k$. The elements of channel noise $N_{k}$ follow the Gaussian distribution with zero mean and standard deviation $\sigma$, where the value of $\sigma$ is determined by the SNR value $\Gamma$. With the received signal, the semantic communication user intercepts the signal to reconstruct the original image with its D-JSCC decoder:
\begin{equation}
    \hat{I}_{k}= D_{\beta}(Y_{I,k},\Gamma), \forall Y_{I,k} \in \mathbb{R}^{k}, \forall \Gamma \in \mathrm{SNR},
\end{equation}
where $D_{\beta}(\cdot)$ represent the semantic-channel decoder of user $k$ and learning parameters $\beta$. Not only does the encoder take the channel condition into account during the encoding process, but the semantic user also considers the channel noise level in its decoding operation. This allows the user to gain knowledge of the range of noise they are dealing with, thereby achieving a better noise elimination process and higher performance.
\subsection{The End-to-end Training Approach and Challenging Issues}
Up to this point, we have clearly defined the D-JSCC framework for the BS encoder and decoder of the semantic communication users, where all the DL-based encoder and decoders are required to be trained together. In a single-user semantic communication scenario, it is common to optimize the DL parameters of the encoder and decoder in an end-to-end manner~\cite{11482529}. Particularly, the image is going through the encoder, channel, and finally the decoder to obtain the reconstruction image $I$. Then, we calculate the mean-square error in every pixel between the original image and the reconstructed one, and then perform backpropagation to update the parameters of the encoder and decoder $\theta$ = \{$\alpha,\beta$\} as presented follows: 
\begin{align}
    &\mathcal{L}_{I,\hat{I}}= \mathrm{MSE}(I,\hat{I}), \forall I \in \mathbb{R}^{n},\\
    & {\theta}^m =  \theta^{m-1} - \eta \nabla \mathcal{L}_{I,\hat{I}},
\end{align}
\begin{figure}[t]
    \centering
    \includegraphics[width=0.5\textwidth]{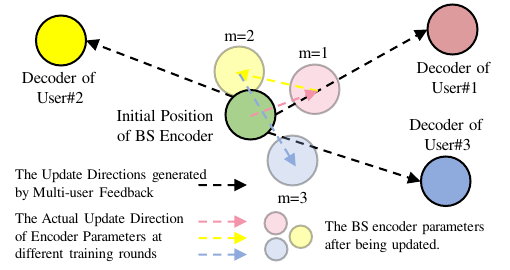}
        \caption{The challenging issue of catastrophic forgetting of the deep learning based encoder for BS when considering multiple users with diverse neural network architectures for the semantic communication system.}
    \label{ParameterUpdateIssue}
\end{figure}
where $m$ indicates the number of training epochs, and $\eta$ is the learning rate of the system. However, this training scenario is limited to a single kind of user, where the semantic encoder updates the parameters by listening to only one feedback. In the proposed semantic communication system, we not only consider scaling up the number of users but also account for their diversity, where the assumption of identical DL-based joint source-channel decoders in previous works no longer holds. Therefore, we employ a diverse range of DL architectures for the semantic communication decoder to reflect the wide range of user kinds. However, this multiple users with diverse DL models is associated with a headache problem, which is referred to as catastrophic forgetting. The problem occurs with the BS encoder when its parameters are updated to serve a group of users, and the updated parameters are not able to encode the data for another group of users. It is essential to notice that the update direction from the decoder of multiple users can be different from each other, and even the probability of conflict among them. As shown in Fig.~\ref{ParameterUpdateIssue}, we illustrate catastrophic forgetting, where the parameters of the BS encoder are trained for three epochs. In the first epoch, the BS encoder is trained in an end-to-end manner with the D-JSCC decoder of user $1$, making progress toward an optimal alignment with that decoder. However, in the second and third epochs, this learned progress deteriorates as the encoder parameters are updated to serve users $2$ and $3$, which is similar to the DL behavior when being trained sequentially for multi-task solving \cite{kirkpatrick2017overcoming}. Therefore, we propose a novel two-stage training framework that not only improves the performance of semantic users but also prevents the forgetting problem, paving the way for the semantic system to scale in terms of the number of users, which is presented in detail in  Section~\ref{Proposed}.

\section{Proposed Two-Stage Training Framework}\label{Proposed}
With the above observation, the forgetting problem arises due to the scattered parameter updates of the BS encoder following the decoder feedback from multiple users. In a nutshell, we aim for the BS encoder to effectively extract and compress semantic features for the DL decoder of multiple semantic users without requiring parameter updates through backpropagation from those decoders. This situation inspires us to come up with the idea of training the DL decoders to align their response to the BS encoder output, where the D-JSCC of the encoder is regarded as the teacher model and fixed. To obtain a BS encoder that can successfully perform semantic extraction from raw data, we first associate it with a DL decoder that possesses a symmetric architecture for training, which we refer to as the \textit{Self-Reflective Training Stage} in Section \ref{SelfReflective}. In the next stage, we train the DL networks of multiple semantic users to align with the output of the BS encoder, which we refer to as the \textit{Adaptive Training Stage for Decoders} in Section \ref{AdaptiveUser}.
\begin{figure}[t]
    \centering
    \includegraphics[width=0.5\textwidth]{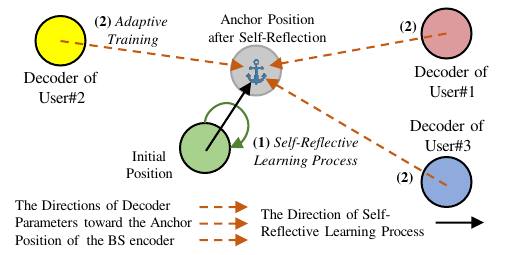}
        \caption{The proposed training framework includes two stages: (1) Self-reflective study of the BS encoder with its symmetric decoder, (2) The adaptive decoders update their model toward the anchor position of the encoder.}
    \label{ProposalParameterUpdates}
\end{figure}

\subsection{Self-Reflective Training Stage of the BS Encoder}\label{SelfReflective}

We first propose an independent decoder that adopts a symmetric architecture to the BS encoder. The symmetric decoder architecture enhances its ability to reverse the encoding process and provides consistent feedback to the encoder. This guarantees stable convergence during training and enhances the BS encoder's semantic extraction and compression. We jointly train the BS encoder and its symmetric decoder, which can be regarded as a form of self-reflective learning applied to the encoder, thereby enabling it to fully utilize its capacity for meaning extraction. The training objective for the self-reflective stage is described as follows:
\begin{equation}
    \argmin_{\alpha, \phi}[\mathcal{L}(I,\hat{I})], \forall I \in \mathbb{R}^{n},
\end{equation}
where $\phi$ denote parameters of independent symmetrical decoder $D_{\phi}$. After completing the training process, we obtain a well-trained BS encoder that can project the raw data into a low-dimensional feature space, which can be transmitted through a noisy wireless channel with minimal distortion. Therefore, we freeze the parameters of the encoder and consider it as the anchor. By solely training the parameters with its mirror, we can avoid the catastrophic forgetting induced by the conflict in gradient descent from the different users.
\begin{figure*}[t]
    \centering
    \includegraphics[width=0.8\textwidth]{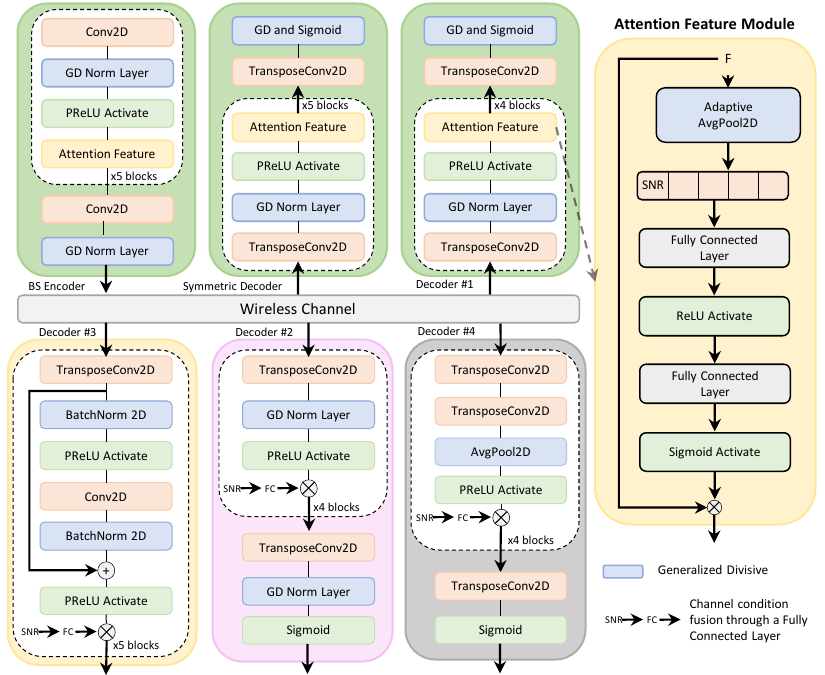}
        \caption{The diversity in deep learning architecture for different semantic communication groups of users, with each group represented by a distinct color.} 
    \label{DLarchitecture}
\end{figure*}
\subsection{Adaptive Training Stage for Multi-users in SemCom}\label{AdaptiveUser}

This subsection details the training procedure for semantic communication users to adapt their learning parameters to the anchor encoder, as well as the DL structure of four representative semantic receivers, as follows.

\subsubsection{Training Process}

With the BS encoder being the anchor, our objective is for the user to successfully interpret the transmitter output, which requires the user's decoder to update its parameters toward the encoder rather than refining the entire network in the conventional training process of semantic communication. Therefore, in our scenario, the DL-based decoders actively align their parameters to adapt to the BS's encoder instead of moving the encoder's network to their desired state. The adaptive training process for user $k$ is denoted as follows.
\begin{align}
        &\hat{I}_{k}= D^{k}_\beta\Big(E_{\alpha}(I,\Gamma)\cdot H_{k}+N_{k},\Big)\\
        & {\beta}^m =  \beta^{m-1} - \eta \nabla \mathcal{L}_{I,\hat{I}_{k}},
\end{align}
where every process is similar to the conventional one, including encoding, transmission, and decoding, except for backpropagation, which stops after the user decoders. A visualization of our two-stage framework is presented in Fig.~\ref{ProposalParameterUpdates}.

\subsubsection{The Architecture of Encoder and the Cross-architecture for DL-based Decoders} We construct four different DL architectures for the joint source-channel decoder of the users to demonstrate the diversity in communication hardware and reflect the real-world scenario. It is worth noticing that each DL architecture can present one group of semantic communication clients. We can increase/decrease the number of layers to simulate the difference in computing capacity for clients in the same group, such as a group of mobile phones: the old generation has fewer layers, and the latest one has more. As shown in Fig.~\ref{DLarchitecture}, we illustrate the representation of DL architectures for BS encoders and decoders. 

\textbf{Attention-based Architecture:} We leverage the attention mechanism to construct the DL-based encoder, the symmetric decoder, and finally the decoder $\textit{1}$. Specifically, the semantic features of the image are extracted by the feature learning group, which includes 2D convolution, Generalized Divisive (GD) normalization, and finally PReLU activation.
\begin{align}
    \mathbf{F}_c &= \mathrm{Conv2D}(\mathbf{x}), \\
    \mathbf{F}_n &= \mathrm{GDN}(\mathbf{F}_c), \\
    \mathbf{F}_a &= \mathrm{PReLU}(\mathbf{F}_n),
\end{align}
where $\mathbf{x}$ denotes the input image and $\mathbf{F}_a$ represents the extracted semantic feature map. The feature map $\mathbf{F}_a$ is then fed into an attention module, where feature importance is evaluated and fused with the SNR to enhance robustness against channel noise. 
Specifically, adaptive average pooling is first applied to aggregate spatial information and capture global context from $\mathbf{F}_a$. The pooled representation is then concatenated with the estimated SNR, followed by a series of fully connected (FC) layers and nonlinear activations to generate attention scores:
\begin{align}
    \mathbf{s}_a &= \mathrm{AvgPool}(\mathbf{F}_a), \\
    \mathbf{s} &= \mathrm{Concat}(\Gamma, \mathbf{s}_a), \\
    \mathbf{w} &= \sigma \big( \mathrm{FC}_2 ( \delta ( \mathrm{FC}_1 (\mathbf{s}) ) ) \big),
\end{align}
where $\Gamma$ denotes the SNR, $\delta(\cdot)$ is the ReLU activation, and $\sigma(\cdot)$ is the sigmoid function. The resulting attention weights $\mathbf{w}$ are then applied to the original feature map to emphasize informative regions and suppress less relevant features.
We employ this DL architecture to construct the BS encoder, its corresponding symmetric decoder, and the decoder for User $1$. The BS encoder and its symmetric decoder share the same architecture, except that each convolutional layer in the encoder is replaced by a transposed convolutional layer in the decoder.

\textbf{Convolution-based Architecture Decoder:} follows a hierarchical upsampling architecture designed to progressively reconstruct the semantic feature representation into the image domain. As shown in Fig~\ref{DLarchitecture}, it comprises multiple stacked blocks, each of which consists of a transposed convolution, followed by a generalized divisive normalization (GDN) layer and a PReLU activation. The transposed convolution layers perform spatial upsampling, gradually increasing the resolution of the feature maps, while the GDN and activation layers enhance nonlinearity and stabilize feature distribution during reconstruction. Finally, unlike the attention module, the channel state information, $\Gamma$ is directly fused with the decoded features at the end of each block via a lightweight FC module, enabling channel-aware feature adaptation.
Let $\mathbf{F}^{(l)}$ denote the input feature at the $l$-th block. The decoding process can be expressed as
\begin{align}
    \tilde{\mathbf{F}}^{(l)} &= \textrm{PReLU} \big( {\textrm{GDN}} ( \mathrm{TConv}(\mathbf{F}^{(l)}) ) \big), \\
    \mathbf{z}^{(l)} &= \mathrm{FC}(\Gamma), \\
    \mathbf{F}^{(l+1)} &= \tilde{\mathbf{F}}^{(l)} \odot \mathbf{z}^{(l)},
\end{align}
where $\textrm{TConv}(\cdot)$ denotes the transposed convolution, and $\odot$ is the element-wise scaling.

\textbf{Residual-based Architecture Decoder:} We adopt a residual learning architecture \cite{he2016deep} as the backbone of Decoder $3$ to facilitate stable training and improve reconstruction performance in deep networks. The residual design mitigates the vanishing gradient problem by introducing identity skip connections, enabling efficient gradient propagation across multiple layers. As illustrated in Fig.~\ref{DLarchitecture}, the decoder is composed of five stacked residual blocks. Each block begins with a transposed convolution layer to perform spatial upsampling, followed by batch normalization and a PReLU activation. The transformed features are then processed by an additional convolution layer and batch normalization. A skip connection directly links the input of the block to its output, and the two paths are combined via element-wise addition, followed by a PReLU activation. This residual mapping allows the network to learn refinement functions rather than complete transformations, thereby improving convergence and representation capability. The SNR is incorporated in the same way as the convolution decoder architecture.
Let $\mathbf{F}^{(l)}$ denote the input to the $l$-th block. The residual mapping is defined as
\begin{align}
    \mathbf{R}^{(l)} &= \mathrm{BN}_2 \big( \mathrm{Conv} ( \delta ( \mathrm{BN}_1 ( \mathrm{TConv}(\mathbf{F}^{(l)}) ) ) ) \big), \\
    \tilde{\mathbf{F}}^{(l)} &= \mathbf{F}^{(l)} + \mathbf{R}^{(l)}, \\
    \mathbf{F}^{(l+1)} &= \delta \big( \tilde{\mathbf{F}}^{(l)} \big) \odot \mathbf{z}^{(l)}.
\end{align}

\textit{VGG-based Architecture:} 
Finally, we adopt a VGG-style architecture~\cite{simonyan2014very} for Decoder~$4$, characterized by a simple and homogeneous stacking of convolutional operations. The decoder is composed of multiple repeated blocks, each designed to progressively refine the semantic representation while maintaining architectural simplicity. Specifically, each block consists of two consecutive TransposeConv2D layers to perform spatial upsampling and feature transformation, followed by an average pooling layer to aggregate local information and stabilize the feature distribution. A PReLU activation is then applied to introduce nonlinearity. Finally, the SNR is incorporated together with the decoded features. This structure is repeated across multiple stages to gradually reconstruct the transmitted signal. Finally, an additional transposed convolution layer is applied to produce the output feature map, followed by a sigmoid activation function to constrain the reconstructed image within a valid range. Compared to the residual-based architecture, this VGG-style design adopts a more straightforward feedforward structure without skip connections.
Let $\mathbf{F}^{(l)}$ denote the input to the $l$-th block. The transformation block can be written as
\begin{align}
    \tilde{\mathbf{F}}^{(l)} &= \delta \big( \mathrm{AvgPool} ( \mathrm{TConv}_2 ( \mathrm{TConv}_1 (\mathbf{F}^{(l)}) ) ) \big), \\
    \mathbf{z}^{(l)} &= \mathrm{FC}(\Gamma), \\
    \mathbf{F}^{(l+1)} &= \tilde{\mathbf{F}}^{(l)} \odot \mathbf{z}^{(l)}.
\end{align}
We provide a complete description of the proposed idea, presented in Algorithm~\ref{alg:two_stage_training}.
\begin{algorithm}[t]
\caption{Anchor-Aided Two-Stage Training Framework}
\label{alg:two_stage_training}
\begin{algorithmic}[1]
\State \textbf{Input:} Dataset $\mathcal{D}=\{I\}$, channel states $\Gamma$, number of users $K$
\State \textbf{Output:} Encoder $E_{\alpha}$ and decoders $\{D^k_{\beta_k}\}_{k=1}^K$

\State \textbf{Stage 1: Self-Reflective Training}
\State Initialize $E_{\alpha}$ and symmetric decoder $D_{\phi}$
\For{each iteration}
    \State $\mathbf{z} \leftarrow E_{\alpha}(I, \Gamma)$
    \State $\tilde{\mathbf{z}} \leftarrow \mathbf{z} \cdot H + N$
    \State $\hat{I} \leftarrow D_{\phi}(\tilde{\mathbf{z}})$
    \State Update $(\alpha, \phi)$ using $\nabla \mathcal{L}(I, \hat{I})$
\EndFor
\State Freeze encoder parameters $\alpha$

\State \textbf{Stage 2: Adaptive Training for Users}
\For{$k = 1$ to $K$}
    \State Initialize decoder $D^k_{\beta_k}$
    \For{each iteration}
        \State $\mathbf{z} \leftarrow E_{\alpha}(I, \Gamma)$
        \State $\tilde{\mathbf{z}}_k \leftarrow \mathbf{z} \cdot H_k + N_k$
        \State $\hat{I}_k \leftarrow D^k_{\beta_k}(\tilde{\mathbf{z}}_k)$
        \State Update $\beta_k$ using $\nabla \mathcal{L}(I, \hat{I}_k)$
    \EndFor
\EndFor
\end{algorithmic}
\end{algorithm}

\subsection{Theoretical Analysis of the Proposed Framework}\label{theorem}
In this subsection, we provide theoretical insights into (i) the origin of catastrophic forgetting in multi-user semantic communication, (ii) the effectiveness of the proposed anchor-based training in eliminating such interference, and (iii) the stability of the self-reflective training stage.

\textit{Theorem 1: Gradient Conflict-Induced Forgetting}

Consider a multi-user semantic communication system with encoder parameters $\alpha$ and user-specific decoder parameters $\{\beta_k\}_{k=1}^K$. Let $\mathcal{L}_k(\alpha, \beta_k)$ denote the reconstruction loss for user $k$. Under a joint training scheme, the encoder is updated as
\begin{equation}
    \alpha^{t+1} = \alpha^t - \eta \sum_{k=1}^K w_k \nabla_\alpha \mathcal{L}_k(\alpha^t, \beta_k^t), \label{updatedrule}
\end{equation}
where $w_k$ are non-negative weights. Then, the loss of a specific user $j$ after one update satisfies
\begin{equation}
    \mathcal{L}_j(\alpha^{t+1}, \beta_j^t)
    \approx
    \mathcal{L}_j(\alpha^t, \beta_j^t)
    - \eta \sum_{k=1}^K w_k 
    \left\langle 
    \nabla_\alpha \mathcal{L}_j, 
    \nabla_\alpha \mathcal{L}_k 
    \right\rangle.
\end{equation}
If there exists $k \neq j$ such that 
$\left\langle \nabla_\alpha \mathcal{L}_j, \nabla_\alpha \mathcal{L}_k \right\rangle < 0$, 
then the update may increase $\mathcal{L}_j$, leading to catastrophic forgetting.

\textit{Proof:}
Using first-order Taylor expansion of loss for user $j$, $\mathcal{L}_j$ around the current encoder parameters $\alpha^t$, we obtain
\begin{equation}
\mathcal{L}_j(\alpha^{t+1}) 
\approx 
\mathcal{L}_j(\alpha^t)
+ \left\langle \nabla_\alpha \mathcal{L}_j, \alpha^{t+1} - \alpha^t \right\rangle.
\end{equation}
Substituting the update rule from Eq.~(\ref{updatedrule}) yields
\begin{equation}
\mathcal{L}_j(\alpha^{t+1})
\approx
\mathcal{L}_j(\alpha^t)
- \eta \sum_{k=1}^K w_k
\left\langle 
\nabla_\alpha \mathcal{L}_j,
\nabla_\alpha \mathcal{L}_k
\right\rangle.
\end{equation}
Therefore, when gradients from different users are negatively aligned, the loss of user $j$ increases after the update. This indicates that optimizing the encoder for one user can adversely affect others, leading to conflicting update directions and ultimately resulting in catastrophic forgetting.

\textit{Theorem 2: Decoupled Optimization with Anchor Encoder}

Let $\alpha^\star$ denote the encoder parameters obtained after the self-reflective training stage and fixed thereafter. Then, the optimization problem for each user decoder becomes
\begin{equation}
    \beta_k^\star = \arg\min_{\beta_k} \mathcal{L}_k(\alpha^\star, \beta_k), \quad k = 1, \dots, K.
\end{equation}
Under this setting, the optimization problems for different users are independent, i.e.,
\begin{equation}
    \frac{\partial \mathcal{L}_i(\alpha^\star, \beta_i)}{\partial \beta_k} = 0, \quad \forall i \neq k.
\end{equation}

\textit{Proof:}
Since $\alpha^\star$ is fixed, the loss function $\mathcal{L}_k(\alpha^\star, \beta_k)$ depends only on $\beta_k$ and not on $\beta_i$ for $i \neq k$. Therefore,
\begin{equation}
    \nabla_{\beta_k} \mathcal{L}_i(\alpha^\star, \beta_i) = 0, \quad \forall i \neq k.
\end{equation}
This implies that the optimization of each decoder parameter $\beta_k$ is independent of others. Hence, no gradient interference occurs across users, and catastrophic forgetting at the encoder is avoided.

\textit{Theorem 3: Monotonic Descent of Self-Reflective Training}

Let $\mathcal{L}_{\text{ref}}(\alpha, \phi)$ denote the reconstruction loss in the self-reflective stage:
\begin{equation}
    \mathcal{L}_{\text{ref}} = \mathbb{E}[\ell(I, \hat{I})],
\end{equation}
where $\hat{I} = D_{\phi}(E_{\alpha}(I, \Gamma))$. Assume that $\mathcal{L}_{\text{ref}}$ is $L$-smooth. Then, for a sufficiently small learning rate $\eta \leq \frac{2}{L}$, gradient descent ensures
\begin{equation}
    \mathcal{L}_{\text{ref}}^{t+1} \leq \mathcal{L}_{\text{ref}}^t.
\end{equation}

\textit{Proof:}
By the $L$-smoothness of $\mathcal{L}_{\text{ref}}$, we have
\begin{equation}
\mathcal{L}_{\text{ref}}(\theta^{t+1})
\leq
\mathcal{L}_{\text{ref}}(\theta^t)
+ \left\langle \nabla \mathcal{L}_{\text{ref}}(\theta^t), \theta^{t+1} - \theta^t \right\rangle
+ \frac{L}{2} \|\theta^{t+1} - \theta^t\|^2,
\end{equation}
where $\theta = (\alpha, \phi)$. Substituting the gradient update
\begin{equation}
\theta^{t+1} = \theta^t - \eta \nabla \mathcal{L}_{\text{ref}}(\theta^t),
\end{equation}
we obtain
\begin{align}
\mathcal{L}_{\text{ref}}(\theta^{t+1})
&\leq
\mathcal{L}_{\text{ref}}(\theta^t)
- \eta \|\nabla \mathcal{L}_{\text{ref}}(\theta^t)\|^2
+ \frac{L \eta^2}{2} \|\nabla \mathcal{L}_{\text{ref}}(\theta^t)\|^2 \\
&=
\mathcal{L}_{\text{ref}}(\theta^t)
- \eta \left(1-\frac{L \eta}{2}\right)
\|\nabla \mathcal{L}_{\text{ref}}(\theta^t)\|^2.
\end{align}
For $\eta \leq \frac{2}{L}$, the right-hand side is non-increasing, yielding
\begin{equation}
\mathcal{L}_{\text{ref}}^{t+1} \leq \mathcal{L}_{\text{ref}}^t.
\end{equation}
Thus, the training is stable and convergent in the sense of monotonic descent.

\begin{figure*}[t]
    \centering
    \includegraphics[width=1\textwidth]{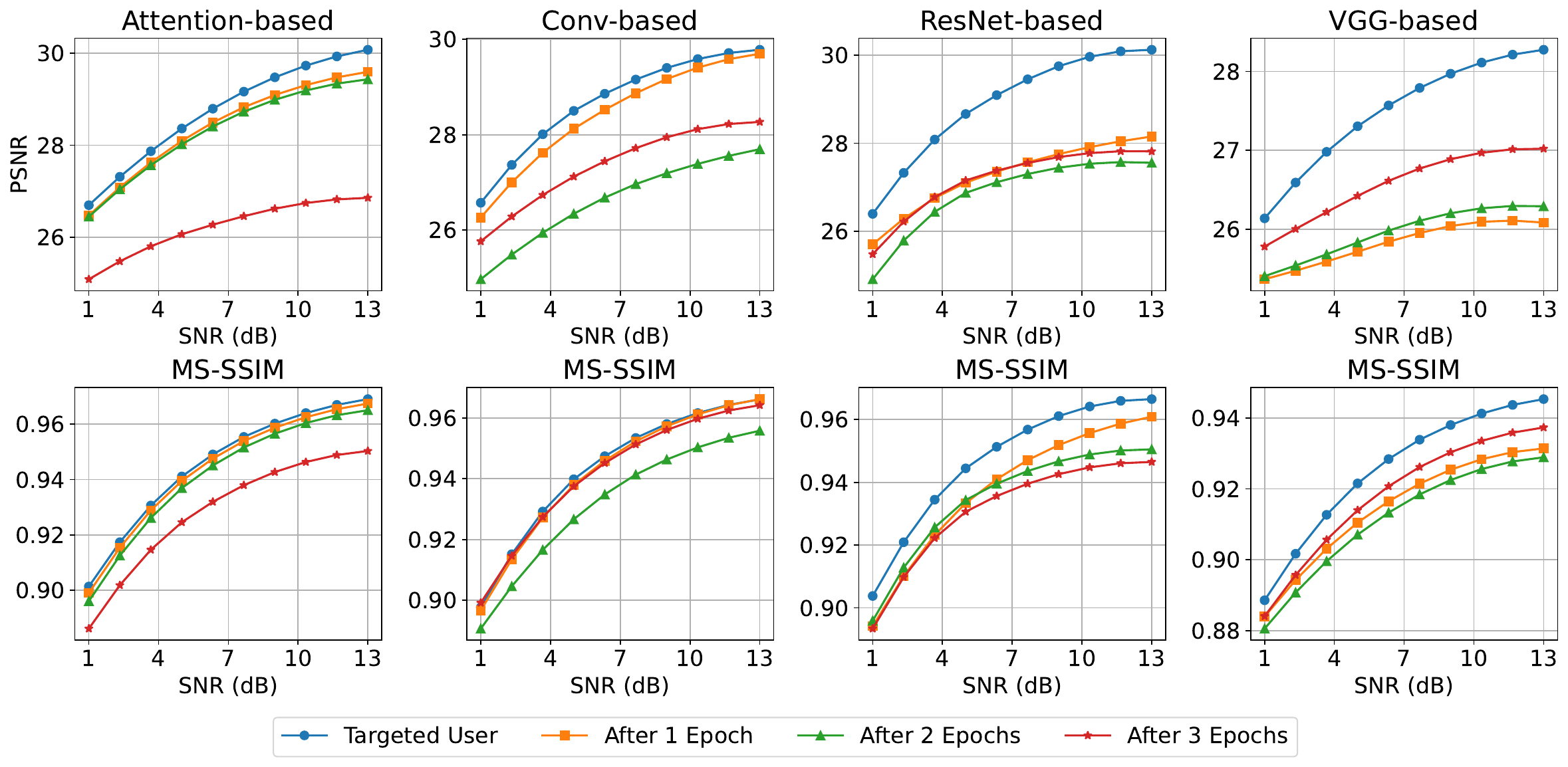}
        \caption{The demonstration for the forgetting problem of deep learning networks in four semantic communication users under Additive White Gaussian Noise.} 
    \label{ForgettingPerforamnce}
\end{figure*}
\section{Performance Evaluation}\label{Performance}

In this section, we conduct various simulations to analyze the effectiveness of the proposed training framework for multiple semantic communication users in the proposed scenario, where different users can be equipped with a unique deep learning decoder. We experiment with our framework in two cases: (1) the wireless channel only experiences the AWGN, and (2) Rayleigh fading along with the AWGN. 

\subsection{Train \& Evaluation Dataset, and Simulation Setup}
We train the proposed multi-user semantic communication model using the DIV2K dataset \cite{agustsson2017ntire}, a well-known dataset renowned for its high-resolution images and widely utilized in semantic communication systems. Then, we evaluate the trained model from different training schemes with the DIV2K validation dataset and the Kodak \cite{franzen1999kodak}. We use the Adam optimization for the training process with the learning $5e^{-4}$, training batch size $40$, the compression rate $1/16$, and the level of AWGN is drawn from the set SNR $\in [1, 4, 7, 10, 13]$ for enhancing the robustness of the system in dealing with a variety of noise and reflect the dynamic of wireless channel.

\subsection{The Training Benchmarks \& Performance Metrics}
To the best of our knowledge, we are the first to tackle the problem of catastrophic forgetting in D-JSCC for multi-user in the semantic communication system, which indicates there is no direct technique for comparison with our training framework. Therefore, we consider two conventional training frameworks as the benchmarks:
\begin{itemize}
    \item \textbf{Iterative Training}: We pair the transmitter encoder with each user's decoder in one training epoch, move to another user's decoder in the next training epoch, and keep repeating the process until every decoder of the user is trained with a pre-defined number of epochs. 
    \item \textbf{Simultaneous Training}: We train the transmitter encoder with all the user's decoders at the same time. The transmitter's parameters are updated by the summation of the losses from all the receivers.
\end{itemize}

Similar to other image reconstruction tasks, we prioritize the two metrics for evaluating the performance of the semantic communication system: 1) the Peak-Signal-Noise-Ratio (PSNR), and 2) the Multi-Scale Structural Similarity Index Metric (MS-SSIM). The higher the value, the better the performance for both metrics, and the maximum value of the MS-SSIM is $1$ while the PSNR metric has no limitation, the smaller the difference in image pixels between two images, the higher the metric value as presented in the following equation:
\begin{equation}
    \textrm{PSNR}= 10\log_{10}\frac{\textrm{MAX}^{2}}{\textrm{MSE}},
    \label{PSNRreverseMSE}
\end{equation}
where $\textrm{MAX}$ is the maximum value of the image pixel in one color channel \cite{bourtsoulatze2019deep}. On the other hand, the MS-SSIM evaluates the reconstruction image at various structural levels, and its equation is presented in detail by the authors in \cite{wang2003multiscale}.

\subsection{Results Analysis}
\subsubsection{The Demonstration of Catastrophic Forgetting}

In Fig.~\ref{ForgettingPerforamnce}, we provide a comprehensive performance of four representative semantic communication users when it is being trained last (targeted) against the performance after the BS encoder is coupled and trained with one other user (After 1 epoch), two other users (After 2 Epochs), and finally three others (After 3 Epochs). Overall, the performance of all semantic users achieves the highest performance in both PSNR and MS-SSIM metrics when the BS encoder is optimized for it. As we expected, if the BS encoder's parameters are trained by the backpropagation of a different decoder, its encoding performance for the previous decoder degrades significantly. One interesting point from this comprehensive simulation result is that not only the number of different matters, but also the difference in the architecture of the user's decoder. To be specific, among the four representative user decoders, the VGG decoder's architecture stands out the most, and its feedback drives the BS encoder parameter far away from the optimal parameters. As we can see, the performance of the Attention-based decoder obtains the highest performance when being the targeted user of the encoder, encountering a slight reduction in performance after 1 epoch (Conv) and 2 epochs (Conv+Resnet). However, its performance dramatically downgrades after being trained with the VGG-based user, specifically from around 29.43 after being trained by two other decoders to 26.86 for the PSNR metric when the signal-to-noise ratio is 13dB.
\begin{figure*}[ht]
    \centering
    \includegraphics[width=1\textwidth]{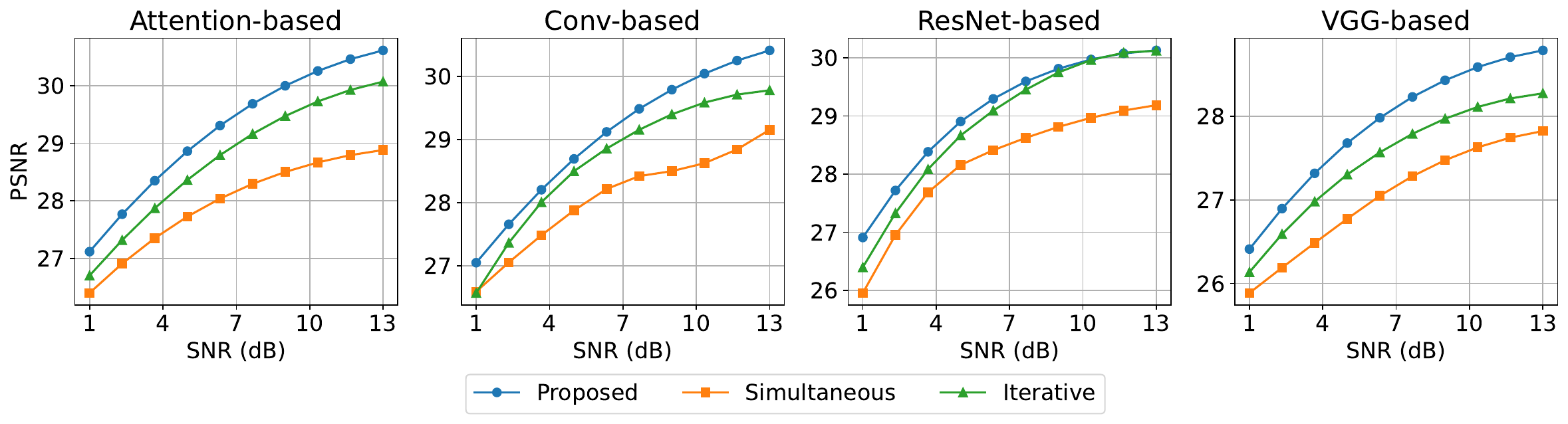}
        \caption{The PSNR comparison among different training approaches for four distinct semantic communication users.} 
    \label{PSNRPerfomanceAWGN}
\end{figure*}

\begin{table}[t]
\renewcommand{\arraystretch}{1.4}
\caption{The comparison in MS-SSIM metric value for different training frameworks under the Additive White Gaussian Noise Channel.}
\centering
\scalebox{0.93}{
\begin{tabular}{|c|c|c|c|c|c|c|}
\hline
AWGN                                                                               &  SNR   &  1dB       & 4dB       &  7 dB      &  10 dB     &  13 dB     \\ \hline
\multirow{3}{*}{\begin{tabular}[c]{@{}c@{}}Attention\\ Decoder\end{tabular}} & Proposed     & \textbf{0.9084} & \textbf{0.9384} & \textbf{0.9560} & \textbf{0.9659} & \textbf{0.9714} \\ \cline{2-7} 
                                                                                   & Iterative    & 0.9014          & 0.9335          & 0.9524          & 0.9631          & 0.9690          \\ \cline{2-7} 
                                                                                   & Simultaneous & 0.8976          & 0.9278          & 0.9450          & 0.9542          & 0.9593          \\ \hline
\multirow{3}{*}{\begin{tabular}[c]{@{}c@{}}Conv\\ Decoder\end{tabular}}      & Proposed     & \textbf{0.9072} & \textbf{0.9369} & \textbf{0.9546} & \textbf{0.9648} & \textbf{0.9705} \\ \cline{2-7} 
                                                                                   & Iterative    & 0.8977          & 0.9322          & 0.9506          & 0.9608          & 0.9661          \\ \cline{2-7} 
                                                                                   & Simultaneous & 0.9006          & 0.9288          & 0.9454          & 0.9548          & 0.9600          \\ \hline
\multirow{3}{*}{\begin{tabular}[c]{@{}c@{}}ResNet\\ Decoder\end{tabular}}    & Proposed     & \textbf{0.9067} & \textbf{0.9413} & \textbf{0.9563} & 0.9632          & 0.9660          \\ \cline{2-7} 
                                                                                   & Iterative    & 0.9038          & 0.9373          & 0.9542          & \textbf{0.9635} & \textbf{0.9665} \\ \cline{2-7} 
                                                                                   & Simultaneous & 0.9007          & 0.9339          & 0.9488          & 0.9570          & 0.9618          \\ \hline
\multirow{3}{*}{\begin{tabular}[c]{@{}c@{}}VGG\\ Decoder\end{tabular}}       & Proposed     & \textbf{0.8901} & \textbf{0.9171} & \textbf{0.9329} & \textbf{0.9418} & \textbf{0.9465} \\ \cline{2-7} 
                                                                                   & Iterative    & 0.8886          & 0.9151          & 0.9313          & 0.9405          & 0.9453          \\ \cline{2-7} 
                                                                                   & Simultaneous & 0.8822          & 0.9095          & 0.9252          & 0.9340          & 0.9386          \\ \hline
\end{tabular}}
\label{MSSSIMAWGN}
\end{table}

\begin{table*}[ht]
\centering
\caption{the comprehensive comparison between the proposed training against the iterative and simultaneous training scenarios under the Rayleigh channel fading}
\renewcommand{\arraystretch}{1.4}
\scalebox{0.95}{ 
\begin{tabular}{|c|c|cc|cc|cc|cc|cc|}
\hline
\multirow{2}{*}{\begin{tabular}[c]{@{}c@{}}Rayleigh\\ Fading Channel\end{tabular}} & \multirow{2}{*}{Frameworks} & \multicolumn{2}{c|}{SNR = 1 dB}        & \multicolumn{2}{c|}{SNR = 4 dB}        & \multicolumn{2}{c|}{SNR = 7 dB}        & \multicolumn{2}{c|}{SNR = 10 dB}       & \multicolumn{2}{c|}{SNR = 13 dB}       \\ \cline{3-12} 
                                                                                   &                             & \multicolumn{1}{c|}{PSNR}    & MS-SSIM & \multicolumn{1}{c|}{PSNR}    & MS-SSIM & \multicolumn{1}{c|}{PSNR}    & MS-SSIM & \multicolumn{1}{c|}{PSNR}    & MS-SSIM & \multicolumn{1}{c|}{PSNR}    & MS-SSIM \\ \hline
\multirow{3}{*}{\begin{tabular}[c]{@{}c@{}}Attention-based\\ Decoder\end{tabular}} & Proposed                    & \multicolumn{1}{c|}{26.0780} & 0.8852  & \multicolumn{1}{c|}{27.0236} & 0.9139  & \multicolumn{1}{c|}{27.6447} & 0.9302  & \multicolumn{1}{c|}{28.0296} & 0.9393  & \multicolumn{1}{c|}{28.2495} & 0.9444  \\ \cline{2-12} 
                                                                                   & Iterative                   & \multicolumn{1}{c|}{25.5375} & 0.8731  & \multicolumn{1}{c|}{26.4341} & 0.9041  & \multicolumn{1}{c|}{27.0173} & 0.9216  & \multicolumn{1}{c|}{27.3628} & 0.9310  & \multicolumn{1}{c|}{27.5389} & 0.9358  \\ \cline{2-12} 
                                                                                   & Simultaneous                & \multicolumn{1}{c|}{25.4916} & 0.8715  & \multicolumn{1}{c|}{26.2931} & 0.9017  & \multicolumn{1}{c|}{26.7889} & 0.9184  & \multicolumn{1}{c|}{27.0826} & 0.9271  & \multicolumn{1}{c|}{27.2380} & 0.9318  \\ \hline
\multirow{3}{*}{\begin{tabular}[c]{@{}c@{}}Conv-based\\ Decoder\end{tabular}}      & Proposed                    & \multicolumn{1}{c|}{25.6425} & 0.8794  & \multicolumn{1}{c|}{26.3626} & 0.9055  & \multicolumn{1}{c|}{26.8536} & 0.9210  & \multicolumn{1}{c|}{27.1292} & 0.9296  & \multicolumn{1}{c|}{27.3229} & 0.9347  \\ \cline{2-12} 
                                                                                   & Iterative                   & \multicolumn{1}{c|}{25.3446} & 0.8700  & \multicolumn{1}{c|}{26.0900} & 0.8996  & \multicolumn{1}{c|}{26.6391} & 0.9169  & \multicolumn{1}{c|}{26.9780} & 0.9267  & \multicolumn{1}{c|}{27.1822} & 0.9318  \\ \cline{2-12} 
                                                                                   & Simultaneous                & \multicolumn{1}{c|}{25.4560} & 0.8732  & \multicolumn{1}{c|}{26.2569} & 0.9020  & \multicolumn{1}{c|}{26.7746} & 0.9181  & \multicolumn{1}{c|}{27.0837} & 0.9268  & \multicolumn{1}{c|}{27.2347} & 0.9313  \\ \hline
\multirow{3}{*}{\begin{tabular}[c]{@{}c@{}}Resnet-based\\ Decoder\end{tabular}}    & Proposed                    & \multicolumn{1}{c|}{25.6765} & 0.8782  & \multicolumn{1}{c|}{26.7301} & 0.9145  & \multicolumn{1}{c|}{27.3056} & 0.9277  & \multicolumn{1}{c|}{27.4807} & 0.9309  & \multicolumn{1}{c|}{27.5160}  & 0.9309  \\ \cline{2-12} 
                                                                                   & Iterative                   & \multicolumn{1}{c|}{25.5007} & 0.8767  & \multicolumn{1}{c|}{26.6080}  & 0.9085  & \multicolumn{1}{c|}{27.2572} & 0.9263  & \multicolumn{1}{c|}{27.5752} & 0.9342  & \multicolumn{1}{c|}{27.6988} & 0.9372  \\ \cline{2-12} 
                                                                                   & Simultaneous                & \multicolumn{1}{c|}{25.4878} & 0.8765  & \multicolumn{1}{c|}{25.7244} & 0.9058  & \multicolumn{1}{c|}{26.1512} & 0.9221  & \multicolumn{1}{c|}{26.4723} & 0.9306  & \multicolumn{1}{c|}{26.7111} & 0.9350  \\ \hline
\multirow{3}{*}{\begin{tabular}[c]{@{}c@{}}VGG-based\\ Decoder\end{tabular}}       & Proposed                    & \multicolumn{1}{c|}{25.1692} & 0.8646  & \multicolumn{1}{c|}{25.8704} & 0.8899  & \multicolumn{1}{c|}{26.2824} & 0.9040  & \multicolumn{1}{c|}{26.5091} & 0.9116  & \multicolumn{1}{c|}{26.6497} & 0.9157  \\ \cline{2-12} 
                                                                                   & Iterative                   & \multicolumn{1}{c|}{25.1374} & 0.8616  & \multicolumn{1}{c|}{25.8096} & 0.8871  & \multicolumn{1}{c|}{26.2821} & 0.9022  & \multicolumn{1}{c|}{26.4972} & 0.9094  & \multicolumn{1}{c|}{26.6365} & 0.9141  \\ \cline{2-12} 
                                                                                   & Simultaneous                & \multicolumn{1}{c|}{24.6598} & 0.8552  & \multicolumn{1}{c|}{25.2163} & 0.8809  & \multicolumn{1}{c|}{25.7058} & 0.8964  & \multicolumn{1}{c|}{26.0380} & 0.9051  & \multicolumn{1}{c|}{26.2442} & 0.9096  \\ \hline
\end{tabular}
}
\label{Rayleigh}
\end{table*}

Furthermore, the simulation results for the convolution-based decoder encounter a similar trend. It is worth noticing that the training sequence is Attention-based $\rightarrow$ Conv-based $\rightarrow$ ResNet-based $\rightarrow$ VGG-based. With this in mind, we observe that the worst performance of the Convolution-based decoder is when the BS encoder's parameters are optimized toward the VGG (after 2 training epochs). It is interesting that the Conv-based decoder's performance actually improves when the encoder is trained using the Attention-based model (from the green line to the red line). This result indicates that the forgetting problem not only depends on the number of training epochs the encoder undergoes, but also on the DL architecture of the decoder. Decoders constructed from a similar block of modules or structures can have optimal points close together, which reduces the catastrophic forgetting effect. However, in reality, it is very difficult for users to obtain synchronization in DL modules or architecture due to the device manufacturer.

\subsubsection{The Effectiveness of the Proposed Framework against Benchmarks} In Fig.~\ref{PSNRPerfomanceAWGN}, we present the performance of our training framework compared to the iterative and simultaneous training cases. In general, our framework achieves the highest performance across all types of decoder users and a wide range of channel qualities. It is important to note that the result of the iterative training case reflects the optimal parameters for each individual user, without considering the catastrophic forgetting problem. Despite the BS encoder’s parameters being frozen during the training of each user’s decoder, our framework still achieves higher performance than the iterative training case, in which the training is conducted in an end-to-end manner. On the other hand, the simultaneous training scenario yields the lowest performance. This can be explained by the way the BS encoder’s parameters are updated. Specifically, simultaneous training aggregates all the losses from individual semantic users and performs a single backpropagation step. As a result, a low-computing decoder with a higher loss tends to dominate the update process, overshadowing the contribution of high-quality decoders with lower losses. Furthermore, the inconsistent update directions among different decoders also contribute to the degraded performance in this scenario.
\begin{figure*}[t]
    \centering
    \includegraphics[width=1\textwidth]{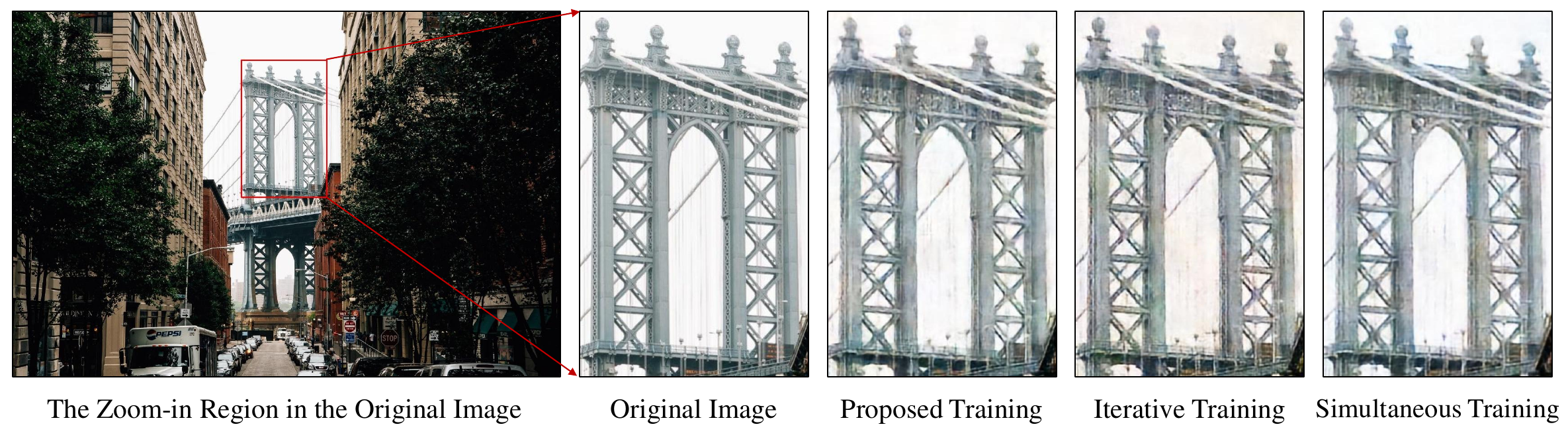}
        \caption{The visualization of the reconstructed images with different training frameworks by the Attention Decoder under the AWGN and the SNR is $1$ dB.} 
    \label{Visual}
\end{figure*}
In Table~\ref{MSSSIMAWGN}, the proposed two-stage training framework outperforms the other two benchmarks in most cases, demonstrating the effectiveness of the anchor decoder. The high performance from the training framework in both metrics indicates that the symmetric has successfully guided the BS's encoder to extract the data meaning and compress it in a meaningful way. After which, in the second stage, each individual learns to interpret the encoded message according to its potential. In particular, the consistent improvement over both iterative and simultaneous cases implies that the encoder learns more generalized and noise-resilient representations, enabling different decoders to interpret the transmitted features more accurately. Furthermore, the performance improvement is higher when the channel contains more noise, which suggests the framework can maintain the semantic meaning of the image under poor channel conditions. For example, the Attention decoder at $1$ dB improves by $0.007$ over iterative and $0.011$ over simultaneous training, and the Conv decoder at $1$ dB improves by $0.010$ over the iterative framework. This improvement may seem small, but considering the maximum value of MS-SSIM is $1$, it represents a significant gap. 

\subsubsection{The Performance Gain for the Rayleigh Fading Channel}

To make our framework more concrete, we have implemented it in a Rayleigh fading channel, where the signal not only suffers from channel noise but also from signal fading due to the blockage of obstacles in the transmission path, as shown in Table~\ref{Rayleigh}. Overall, the performance of the semantic communication models degrades for the decoders in all training frameworks compared to the AWGN channel due to the effects of Rayleigh fading from the environment. Among the decoders, the attention-based decoder achieves the highest performance in both PSNR and MS-SSIM metrics, followed by the ResNet and convolution-based decoders, with the VGG model showing the lowest performance. 

In detail, the Attention-based decoder in our two-stage framework achieves $26.0780$ in PSNR and $0.8852$ in MS-SSIM metrics at an SNR of $1$ dB, compared with $25.5375$ and $0.8731$ under the iterative scenario, and $25.4916$ and $0.8715$ under the simultaneous scenario. The performance gap is approximately $0.5$ value for the PSNR and $0.012$ for the MS-SSIM. When the wireless channel quality improves from $1$ dB to $13$ dB, the performance of the semantic communication users improves, to be precise, $2.1715$ and $0.0592$ for the proposed framework, $2.0014$ and $0.0627$ for the iterative one, $1.7476$ and $0.0603$ under the simultaneous case. These results are from the attention-based decoder, but similar trends can be observed in all other users, for example, an improvement from $25.1692$ and $0.86460$ to $26.6497$ and $0.9157$ of the VGG-based decoder under the two-stage training framework.

\subsubsection{Visualization differences in reconstructed image among the training frameworks} In Fig.~\ref{Visual}, we provide an example of the reconstructed image from our two-stage training framework versus the iterative and simultaneous cases, which are the results from the attention-based decoder at a noise level of 1 dB. Overall, all the images look similar to the original one, demonstrating the capability of the semantic communication framework. However, the differences become noticeable when zooming in, as shown in the figure. Among the images, our proposed method maintains consistency in pixel colors and most closely resembles the original, while the simultaneous framework exhibits some blue regions in its reconstructed image. This comparison in visualization once again confirms the effectiveness of our two-stage training framework.

\section{Conclusion}\label{Conclusion}

In this paper, we propose a downlink communication framework from a single base station to multiple users using semantic communication. In our scenario, each user is equipped with a unique deep learning model for joint source–channel coding as the semantic decoder, reflecting the asynchronous nature of device resources in real-world environments. This asynchrony in the DL architecture introduces the catastrophic forgetting problem in neural networks, as the base station encoder must be trained to serve a wide range of users. To address this issue, we proposed a novel training framework in which the base station encoder is first trained with a symmetric decoder in an end-to-end manner, representing a self-reflective learning process. The compatibility in the deep learning model between the encoder and its symmetric decoder facilitates the optimization of the base station encoder parameters. After this stage, the base station encoder has successfully encoded the semantic meaning of the image and therefore its parameters are frozen, and leveraged to train multiple semantic users to adapt to its output. This frozen encoder acts as an anchor, and all other semantic users update their parameters to align with it. Finally, we conducted extensive simulations to demonstrate both the effects of catastrophic forgetting in semantic communication and the effectiveness of the proposed framework compared with iterative and simultaneous training under two types of channels: Additive White Gaussian Noise and Rayleigh fading. Our proposed framework achieves higher performance compared to existing training schemes across almost all levels of channel noise in both channel models.

\appendices

\ifCLASSOPTIONcaptionsoff
  \newpage
\fi

\bibliographystyle{IEEEtran}
\bibliography{mybib}

\end{document}